\documentclass[10pt,conference,margin=1in]{IEEEtran}
\IEEEoverridecommandlockouts

\ifCLASSINFOpdf
  \usepackage[pdftex]{graphicx}
\else
\fi
\usepackage{cite}
\usepackage{graphicx}
\usepackage{textcomp}
\usepackage{xcolor}
\def\BibTeX{{\rm B\kern-.05em{\sc i\kern-.025em b}\kern-.08em
    T\kern-.1667em\lower.7ex\hbox{E}\kern-.125emX}}

\usepackage[T1]{fontenc}
\usepackage{siunitx}
\usepackage{cite}
\usepackage[cmex10]{amsmath}
\usepackage{amssymb}
\usepackage{makecell}
\usepackage{xfrac}
\usepackage{boldline}
\usepackage{adjustbox}
\usepackage{caption}
\usepackage{tikz}
\usetikzlibrary{shapes,arrows,fit,positioning,shadows,calc}
\usetikzlibrary{plotmarks}
\usetikzlibrary{decorations.pathreplacing}
\usetikzlibrary{patterns}
\usepackage{algorithm}
\usepackage{algpseudocode}
\algnewcommand{\Initialization}[1]{%
  \State \textbf{initialization:}
  \Statex \hspace*{\algorithmicindent}\parbox[t]{.8\linewidth}{\raggedright #1}
}
\usetikzlibrary{automata}
\usepackage{xcolor}
\usetikzlibrary{spy}

\usepackage{pgfplots}
\pgfplotsset{compat=newest}

\usepackage{graphicx}
\usepackage{ragged2e}
\usepackage{hhline}

\newcommand{\pr}[1]{\ensuremath{\left[#1\right]}}
\newcommand{\pc}[1]{\ensuremath{\left(#1\right)}}
\newcommand{\chav}[1]{\ensuremath{\left\{#1\right\}}}
\newcommand{\PM}[1]{\ensuremath{\left|#1\right|}}

\definecolor{b}{rgb}{0, 0, 1}
\definecolor{r}{rgb}{1, 0, 0}
\definecolor{r2}{rgb}{0,0,0}
\definecolor{r3}{rgb}{0,0,0}

\definecolor{dark_green}{rgb}{0, 0.33, 0.13}
\definecolor{naplesyellow}{rgb}{0.99, 0.93, 0.0}
\definecolor{aureolin}{rgb}{1, 0.8, 0}
\definecolor{purple}{rgb}{0.4940 0.1840 0.5560}

\begin{document}
\title{Symbol-Error Probability Constrained Power Minimization for Reconfigurable Intelligent Surfaces-based Passive Transmitter}


\author{\IEEEauthorblockN{Erico~S.~P.~Lopes and Lukas~T.~N.~Landau}
\IEEEauthorblockA{\textit{Pontifical Catholic University of Rio de Janeiro} \\
	Rio de Janeiro, Brazil 22453-900 \\
erico@aluno.puc-rio.br;landau@puc-rio.br}
\vspace{-1.5em}
\thanks{This work was supported by Amazônia Azul Tecnologias de Defesa S.A., CAPES, FAPERJ and FAPESP.}
}

\maketitle

\begin{abstract}
This study considers a virtual multiuser multiple-input multiple-output system with PSK modulation realized via the reconfigurable intelligent surface-based passive transmitter setup. Under this framework, the study derives the formulation for the union-bound symbol-error probability, which is an upper bound on the actual symbol-error probability. Based on this, a symbol-level precoding power minimization problem under the condition that the union-bound symbol-error probability is below a given requirement is proposed. The problem is formulated as a constrained optimization on an oblique manifold, and solved via a bisection method. The method consists of successively optimizing transmit power while evaluating the feasibility of the union-bound symbol-error probability requisite by solving, via the Riemannian conjugate gradient algorithm, an auxiliary problem dependent only on the reflection coefficients of the reconfigurable intelligent surface elements. Numerical results demonstrate the effectiveness of the proposed approach in minimizing the transmit power for different symbol-error probability requirements.
\end{abstract}


\section{Introduction}
\label{sec:intro}

To enable the new key applications of the next generation of wireless communications, attainment to strict quality-of-service (QoS) requisites, such as low latency, high reliability, and, high data rate is required \cite{Ris_6G,6G_100GHz}. As stated in \cite{6G_Future_Directions}, the future wireless generation will require an improvement in the data rate of factor 100 for the uplink and factor 50 for the downlink while achieving 10 thousand times higher reliability when compared to 5G. Massive multiuser multiple-input multiple-output (MU-MIMO) systems are considered as a promising technique and are expected to be a key technology for attaining these requirements \cite{6G_Future_Directions}. Yet, high hardware costs and increased energy consumption yield bottlenecks for the practical implementation of such technologies.

One approach to realizing low-cost energy-efficient massive MIMO is the utilization of reconfigurable intelligent surfaces (RIS). RIS are two-dimensional surfaces with many reconfigurable passive reflecting elements that can independently adjust their reflection coefficient in a real-time programmable manner. As first proposed in \cite{Basar_access2019}, RIS can be utilized as a passive transmitter by changing the parameters of the reflecting elements to modulate and transmit information symbols by exploiting an unmodulated carrier signal generated by a nearby radio-frequency (RF) signal generator. With this, RIS-based passive transmitters realize virtual MIMO systems with a single RF chain and cost-effective reflecting elements, which benefits the implementation of massive MIMO with reduced hardware complexity and increased energy efficiency. 

The RIS-based passive transmission scheme yields similar mechanism as in symbol-level precoding (SLP) \cite{General_MMDDT_BB,lopes2021discrete,lopes_wcl2022,lopes_tcom2023,masouros_twc2018,MSM_precoder,CVX-CIO}, which, by varying the precoder in a symbol-by-symbol way, can exploit multiuser interference to achieve high-performance systems. Different works, consider RIS-based passive transmission schemes. In \cite{li2022reconfigurable} the authors jointly optimize the total power reflected from the RIS and the power allocation fraction assigned to each user. In \cite{liu2021intelligent} a power minimization problem is considered under the condition that the minimum distance to the decisions threshold (MDDT) of each user's noiseless received symbol is greater or equal to a given requirement.

Following the path of \cite{liu2021intelligent}, this study proposes a power minimization problem under QoS requisites. Yet, different than in \cite{liu2021intelligent}, the considered requisite is union-bound symbol-error probability (SEP), first proposed for MU-MIMO downlink in \cite{lopes_globecom2022}. As discussed in \cite{lopes_globecom2022}, the union-bound SEP is an upper bound on the SEP, with this, restraining it to a maximum value guarantees the attainment of the system's SEP requisite. 
The problem is formulated as a constrained optimization on an oblique manifold, and solved via a bisection method (BM). The proposed BM successively adjusts the transmit power while evaluating the feasibility of the union-bound SEP requisite by solving, via the Riemannian conjugate gradient (RCG) algorithm, an auxiliary problem dependent only on the coefficients of the RIS reflecting elements. Numerical results underline the effectiveness of the proposed method in minimizing the transmit power for different SEP requirements. 

The remainder of this paper is organized as follows: Section~\ref{sec:system_model} describes the system model. Section~\ref{sec:precoding} formulates the power minimization problem under QoS constraints. Section~\ref{sec:bisection} describes the algorithms utilized for solving it. Section~\ref{sec:numerical_results} discusses numerical results and Section \ref{sec:conclusion} gives the conclusions.
Regarding the notation for a given matrix $\boldsymbol{A}$, $\pr{\boldsymbol{A}}_{i,j}$ denotes the element of the $i$-th row and $j$-th column and $\pr{\boldsymbol{A}}_{(i,:)}$ denotes $i$-th row of $\boldsymbol{A}$. For the given vectors $\boldsymbol{a}$ and $\boldsymbol{b}$, $\text{P}(\boldsymbol{a}=\boldsymbol{b})$ denotes the probability of the event $\boldsymbol{a}=\boldsymbol{b}$.

\section{System Model}
\label{sec:system_model}
\begin{figure}
\captionsetup{justification=centering}
\centering
\tikzset{every picture/.style={line width=0.75pt}} 

\begin{tikzpicture}[x=0.4pt,y=0.4pt,yscale=-1,xscale=1]

\draw  [color={rgb, 255:red, 155; green, 155; blue, 155 }  ,draw opacity=1 ][fill={rgb, 255:red, 155; green, 155; blue, 155 }  ,fill opacity=1 ] (237.77,227.25) -- (178.61,222.23) -- (175.26,215.47) -- (207.1,165.35) -- cycle ;
\draw  [color={rgb, 255:red, 155; green, 155; blue, 155 }  ,draw opacity=1 ][fill={rgb, 255:red, 155; green, 155; blue, 155 }  ,fill opacity=1 ] (161.37,223.03) -- (175,215.61) -- (178.61,222.23) -- (164.97,229.65) -- cycle ;
\draw    (163.13,226.28) -- (144.8,236.72) ;
\draw   (126.4,245.91) .. controls (123.86,240.83) and (125.92,234.65) .. (131,232.11) .. controls (136.08,229.57) and (142.26,231.64) .. (144.8,236.72) .. controls (147.34,241.8) and (145.28,247.98) .. (140.2,250.52) .. controls (135.11,253.06) and (128.94,251) .. (126.4,245.91) -- cycle ;
\draw   (129.41,244.17) .. controls (136.19,231.05) and (134.03,252.78) .. (140.8,239.66) ;
\draw  [color={rgb, 255:red, 74; green, 144; blue, 226 }  ,draw opacity=1 ][fill={rgb, 255:red, 74; green, 144; blue, 226 }  ,fill opacity=1 ] (333.32,62) -- (563.68,62) -- (563.68,193.21) -- (333.32,193.21) -- cycle ;
\draw  [color={rgb, 255:red, 255; green, 255; blue, 255 }  ,draw opacity=1 ][fill={rgb, 255:red, 255; green, 255; blue, 255 }  ,fill opacity=1 ] (354.26,80.74) -- (375.21,80.74) -- (375.21,99.49) -- (354.26,99.49) -- cycle ;
\draw  [color={rgb, 255:red, 255; green, 255; blue, 255 }  ,draw opacity=1 ][fill={rgb, 255:red, 255; green, 255; blue, 255 }  ,fill opacity=1 ] (396.15,80.74) -- (417.09,80.74) -- (417.09,99.49) -- (396.15,99.49) -- cycle ;
\draw  [color={rgb, 255:red, 255; green, 255; blue, 255 }  ,draw opacity=1 ][fill={rgb, 255:red, 255; green, 255; blue, 255 }  ,fill opacity=1 ] (438.03,80.74) -- (458.97,80.74) -- (458.97,99.49) -- (438.03,99.49) -- cycle ;
\draw  [color={rgb, 255:red, 255; green, 255; blue, 255 }  ,draw opacity=1 ][fill={rgb, 255:red, 255; green, 255; blue, 255 }  ,fill opacity=1 ] (479.92,80.74) -- (500.86,80.74) -- (500.86,99.49) -- (479.92,99.49) -- cycle ;
\draw  [color={rgb, 255:red, 255; green, 255; blue, 255 }  ,draw opacity=1 ][fill={rgb, 255:red, 255; green, 255; blue, 255 }  ,fill opacity=1 ] (521.8,80.74) -- (542.74,80.74) -- (542.74,99.49) -- (521.8,99.49) -- cycle ;
\draw  [color={rgb, 255:red, 255; green, 255; blue, 255 }  ,draw opacity=1 ][fill={rgb, 255:red, 255; green, 255; blue, 255 }  ,fill opacity=1 ] (354.26,118.23) -- (375.21,118.23) -- (375.21,136.98) -- (354.26,136.98) -- cycle ;
\draw  [color={rgb, 255:red, 255; green, 255; blue, 255 }  ,draw opacity=1 ][fill={rgb, 255:red, 255; green, 255; blue, 255 }  ,fill opacity=1 ] (396.15,118.23) -- (417.09,118.23) -- (417.09,136.98) -- (396.15,136.98) -- cycle ;
\draw  [color={rgb, 255:red, 255; green, 255; blue, 255 }  ,draw opacity=1 ][fill={rgb, 255:red, 255; green, 255; blue, 255 }  ,fill opacity=1 ] (438.03,118.23) -- (458.97,118.23) -- (458.97,136.98) -- (438.03,136.98) -- cycle ;
\draw  [color={rgb, 255:red, 255; green, 255; blue, 255 }  ,draw opacity=1 ][fill={rgb, 255:red, 255; green, 255; blue, 255 }  ,fill opacity=1 ] (479.92,118.23) -- (500.86,118.23) -- (500.86,136.98) -- (479.92,136.98) -- cycle ;
\draw  [color={rgb, 255:red, 255; green, 255; blue, 255 }  ,draw opacity=1 ][fill={rgb, 255:red, 255; green, 255; blue, 255 }  ,fill opacity=1 ] (521.8,118.23) -- (542.74,118.23) -- (542.74,136.98) -- (521.8,136.98) -- cycle ;
\draw  [color={rgb, 255:red, 255; green, 255; blue, 255 }  ,draw opacity=1 ][fill={rgb, 255:red, 255; green, 255; blue, 255 }  ,fill opacity=1 ] (354.26,155.72) -- (375.21,155.72) -- (375.21,174.46) -- (354.26,174.46) -- cycle ;
\draw  [color={rgb, 255:red, 255; green, 255; blue, 255 }  ,draw opacity=1 ][fill={rgb, 255:red, 255; green, 255; blue, 255 }  ,fill opacity=1 ] (396.15,155.72) -- (417.09,155.72) -- (417.09,174.46) -- (396.15,174.46) -- cycle ;
\draw  [color={rgb, 255:red, 255; green, 255; blue, 255 }  ,draw opacity=1 ][fill={rgb, 255:red, 255; green, 255; blue, 255 }  ,fill opacity=1 ] (438.03,155.72) -- (458.97,155.72) -- (458.97,174.46) -- (438.03,174.46) -- cycle ;
\draw  [color={rgb, 255:red, 255; green, 255; blue, 255 }  ,draw opacity=1 ][fill={rgb, 255:red, 255; green, 255; blue, 255 }  ,fill opacity=1 ] (479.92,155.72) -- (500.86,155.72) -- (500.86,174.46) -- (479.92,174.46) -- cycle ;
\draw  [color={rgb, 255:red, 255; green, 255; blue, 255 }  ,draw opacity=1 ][fill={rgb, 255:red, 255; green, 255; blue, 255 }  ,fill opacity=1 ] (521.8,155.72) -- (542.74,155.72) -- (542.74,174.46) -- (521.8,174.46) -- cycle ;
\draw    (222,197.65) -- (435.22,119.27) ;
\draw [shift={(438.03,118.23)}, rotate = 159.82] [fill={rgb, 255:red, 0; green, 0; blue, 0 }  ][line width=0.08]  [draw opacity=0] (8.93,-4.29) -- (0,0) -- (8.93,4.29) -- cycle    ;
\draw    (429.21,139.94) -- (334.6,340.45) ;
\draw [shift={(333.32,343.16)}, rotate = 295.26] [fill={rgb, 255:red, 0; green, 0; blue, 0 }  ][line width=0.08]  [draw opacity=0] (8.93,-4.29) -- (0,0) -- (8.93,4.29) -- cycle    ;
\draw    (473.3,139.94) -- (583.18,340.53) ;
\draw [shift={(584.62,343.16)}, rotate = 241.29] [fill={rgb, 255:red, 0; green, 0; blue, 0 }  ][line width=0.08]  [draw opacity=0] (8.93,-4.29) -- (0,0) -- (8.93,4.29) -- cycle    ;
\draw    (449.05,145.85) -- (458.84,358.91) ;
\draw [shift={(458.97,361.9)}, rotate = 267.37] [fill={rgb, 255:red, 0; green, 0; blue, 0 }  ][line width=0.08]  [draw opacity=0] (8.93,-4.29) -- (0,0) -- (8.93,4.29) -- cycle    ;
\draw   (312.38,359.93) -- (327.81,359.93) -- (327.81,380.65) -- (312.38,380.65) -- cycle ;
\draw    (312.38,343.16) -- (312.38,359.93) ;
\draw   (443.54,378.67) -- (458.97,378.67) -- (458.97,399.39) -- (443.54,399.39) -- cycle ;
\draw    (443.54,361.9) -- (443.54,378.67) ;
\draw   (569.19,359.93) -- (584.62,359.93) -- (584.62,380.65) -- (569.19,380.65) -- cycle ;
\draw    (569.19,343.16) -- (569.19,359.93) ;
\draw   (314.59,362.89) -- (325.61,362.89) -- (325.61,368.81) -- (314.59,368.81) -- cycle ;
\draw   (445.75,381.63) -- (456.77,381.63) -- (456.77,387.55) -- (445.75,387.55) -- cycle ;
\draw   (571.4,362.89) -- (582.42,362.89) -- (582.42,368.81) -- (571.4,368.81) -- cycle ;
\draw    (136,252) -- (136,258) ;
\draw    (130,258) -- (142,258) ;
\draw    (132,260) -- (140,260) ;
\draw    (134,262) -- (138,262) ;
\draw  [fill={rgb, 255:red, 184; green, 233; blue, 134 }  ,fill opacity=1 ] (599.8,84.69) -- (654,84.69) -- (654,124.15) -- (599.8,124.15) -- cycle ;
\draw  [fill={rgb, 255:red, 177; green, 3; blue, 3 }  ,fill opacity=1 ] (606.18,145.85) -- (647.62,145.85) -- (647.62,173.48) -- (606.18,173.48) -- cycle ;
\draw    (626.9,124.15) -- (626.9,145.85) ;
\draw    (563.93,105) -- (599.8,105) ;

\draw (503.29,363.18) node  [scale=0.8]    {$\dotsc $};
\draw (320.84,400.41) node  [scale=0.8]   {User 1};
\draw (451.51,419.14) node  [scale=0.8]   {User 2};
\draw (586.22,400.46) node  [scale=0.8]   {User $\displaystyle K$};
\draw (290.31,150.4) node  [scale=0.8]    {$\boldsymbol{h}_{g}$};
\draw (353.37,246.75) node  [scale=0.8]    {$\boldsymbol{h}_{u_{1}}$};
\draw (432.13,258.58) node  [scale=0.8]    {$\boldsymbol{h}_{u_{2}}$};
\draw (562.48,256.75) node  [scale=0.8]    {$\boldsymbol{h}_{u_{K}}$};
\draw (488.96,262.56) node  [scale=0.8]    {$\dotsc $};
\draw (137,162) node [scale=0.8] {\begin{minipage}[lt]{46.92pt}\setlength\topsep{0pt}
\begin{center}
RF \\Generator
\end{center}
\end{minipage}};
\draw (626.5,105) node  [scale=0.48]   {Controller};
\draw (626.5,160.67) node  [scale=0.48]   {Source};

\end{tikzpicture}
\caption{Multiuser MIMO downlink via passive RIS reflection}
\label{fig:system_model}       
\end{figure}
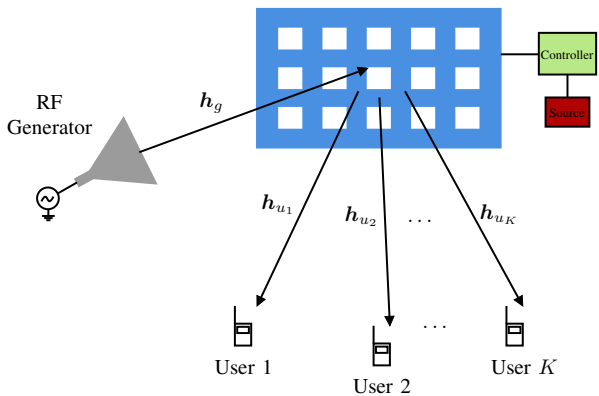
The system model, shown in Fig.~\ref{fig:system_model}, consists of an RF generator illuminating a RIS with $N$ reflecting elements that serve $K$ single antenna users. A symbol-level transmission is considered with the symbols generated by a memoryless source connected to the RIS controller. The data symbol of the $k$-th user is denoted as $s_k$, such that for all $k\in\mathcal{K}=\chav{1,\hdots,K}$, $s_k\in \mathcal{S}$, where $\mathcal{S}$ represents all possible symbols of a $\alpha_{s}$-PSK modulation. The symbols of all users are described in a stacked vector notation as $\boldsymbol{s}=[{s}_1,\ldots,{s}_K]^T$. Based on $\boldsymbol{s}$ the controller determines the phase shift vector $\boldsymbol{\theta}=\pr{\theta_1,\hdots,\theta_N}^T$, where $\PM{\pr{\boldsymbol{\theta}}_n}^2=1, \ \forall n\in \mathcal{N}=\chav{1,\hdots,N}$. 
Given the availability of channel estimation approaches \cite{Qingqing_irstutorial,zhang_cest_twc,Araújo_cest_jstsp,Zhang_cest_jsac}, perfect channel state information at the RIS is considered. 

With this, the received signal of the $k$-th user $z_k$, for all $k\in \mathcal{K}$, reads as ${z}_k= \sqrt{P}\boldsymbol{h}_{k}^H\boldsymbol{\theta}+{w}_k$, 
with $P$ being the transmit power of the RF generator, ${w}_k\sim \mathcal{CN}({0},\sigma_w^2)$ representing additive white Gaussian noise, and, $\boldsymbol{h}_k=\boldsymbol{h}_{u_k}^H\text{diag}\pc{\boldsymbol{h}_{g}}$, being the effective channel, where $\boldsymbol{h}_{g}$ is the channel between the RF generator and the RIS and $\boldsymbol{h}_{u_k}$ is the channel between the RIS the $k$-th user. 

Each $z_k$ is hard detected based on the decision region it belongs. The decision region of $s_i$, termed $\mathcal{S}_i$, is the set of points closer to $s_i$ than all other valid candidates for detection. This implies that $z_k$ is detected as $s_i$ if $z_k \in \mathcal{S}_i$. For PSK the decision regions are circle sectors with infinite radius and angle of $2\phi$, where $\phi=\sfrac{\pi}{\alpha_s}$. Finally, the detected symbol vector is written as $\hat{\boldsymbol{s}}=\pr{\hat{s}_1, \hdots, \hat{s}_K}$.
\section{Power Minimization Problem Formulation}
\label{sec:precoding}
\begin{figure}[t] 
\centering
\tikzset{every picture/.style={line width=0.75pt}} 

\begin{tikzpicture}[x=0.25pt,y=0.25pt,yscale=-1,xscale=1]

\draw    (236,421) -- (561,95) ;
\draw    (235.13,95.41) -- (559.13,418.41) ;
\draw  (152,257.41) -- (644,257.41)(398.13,23) -- (398.13,493) (637,252.41) -- (644,257.41) -- (637,262.41) (393.13,30) -- (398.13,23) -- (403.13,30)  ;
\draw  [color={rgb, 255:red, 65; green, 117; blue, 5 }  ,draw opacity=1 ][fill={rgb, 255:red, 65; green, 117; blue, 5 }  ,fill opacity=1 ] (438.75,142) -- (442.42,142) -- (442.42,144.75) -- (445.17,144.75) -- (445.17,148.58) -- (442.42,148.58) -- (442.42,151.33) -- (438.75,151.33) -- (438.75,148.58) -- (436,148.58) -- (436,144.75) -- (438.75,144.75) -- cycle ;
\draw  [color={rgb, 255:red, 255; green, 0; blue, 0 }  ,draw opacity=1 ][fill={rgb, 255:red, 252; green, 0; blue, 0 }  ,fill opacity=1 ] (507.58,210) -- (511.25,210) -- (511.25,212.75) -- (514,212.75) -- (514,216.58) -- (511.25,216.58) -- (511.25,219.33) -- (507.58,219.33) -- (507.58,216.58) -- (504.83,216.58) -- (504.83,212.75) -- (507.58,212.75) -- cycle ;
\draw  [color={rgb, 255:red, 65; green, 117; blue, 5 }  ,draw opacity=1 ][fill={rgb, 255:red, 65; green, 117; blue, 5 }  ,fill opacity=1 ] (506.75,296.67) -- (510.42,296.67) -- (510.42,299.42) -- (513.17,299.42) -- (513.17,303.25) -- (510.42,303.25) -- (510.42,306) -- (506.75,306) -- (506.75,303.25) -- (504,303.25) -- (504,299.42) -- (506.75,299.42) -- cycle ;
\draw  [color={rgb, 255:red, 0; green, 0; blue, 0 }  ,draw opacity=1 ][fill={rgb, 255:red, 0; green, 0; blue, 0 }  ,fill opacity=1 ] (541.58,182) -- (545.25,182) -- (545.25,184.75) -- (548,184.75) -- (548,188.58) -- (545.25,188.58) -- (545.25,191.33) -- (541.58,191.33) -- (541.58,188.58) -- (538.83,188.58) -- (538.83,184.75) -- (541.58,184.75) -- cycle ;
\draw    (506.47,152.36) -- (541.95,185.31) ;
\draw [shift={(543.42,186.67)}, rotate = 222.87] [color={rgb, 255:red, 0; green, 0; blue, 0 }  ][line width=0.75]    (10.93,-3.29) .. controls (6.95,-1.4) and (3.31,-0.3) .. (0,0) .. controls (3.31,0.3) and (6.95,1.4) .. (10.93,3.29)   ;
\draw [shift={(505,151)}, rotate = 42.87] [color={rgb, 255:red, 0; green, 0; blue, 0 }  ][line width=0.75]    (10.93,-3.29) .. controls (6.95,-1.4) and (3.31,-0.3) .. (0,0) .. controls (3.31,0.3) and (6.95,1.4) .. (10.93,3.29)   ;
\draw    (543.43,188.67) -- (543.98,256) ;
\draw [shift={(544,258)}, rotate = 269.53] [color={rgb, 255:red, 0; green, 0; blue, 0 }  ][line width=0.75]    (10.93,-3.29) .. controls (6.95,-1.4) and (3.31,-0.3) .. (0,0) .. controls (3.31,0.3) and (6.95,1.4) .. (10.93,3.29)   ;
\draw [shift={(543.42,186.67)}, rotate = 89.53] [color={rgb, 255:red, 0; green, 0; blue, 0 }  ][line width=0.75]    (10.93,-3.29) .. controls (6.95,-1.4) and (3.31,-0.3) .. (0,0) .. controls (3.31,0.3) and (6.95,1.4) .. (10.93,3.29)   ;
\draw  [draw opacity=0][fill={rgb, 255:red, 126; green, 211; blue, 33 }  ,fill opacity=0.34 ] (662.28,256.12) .. controls (662.28,256.12) and (662.28,256.12) .. (662.28,256.12) .. controls (662.28,256.12) and (662.28,256.12) .. (662.28,256.12) .. controls (662.28,389.33) and (544.35,497.32) .. (398.89,497.32) .. controls (330.55,497.32) and (268.29,473.49) .. (221.48,434.4) -- (398.89,256.12) -- cycle ;
\draw  [draw opacity=0][fill={rgb, 255:red, 27; green, 101; blue, 196 }  ,fill opacity=0.38 ] (135.11,258) .. controls (135.11,258) and (135.11,258) .. (135.11,258) .. controls (135.11,258) and (135.11,258) .. (135.11,258) .. controls (135.11,124.79) and (253.03,16.8) .. (398.5,16.8) .. controls (466.84,16.8) and (529.1,40.63) .. (575.91,79.72) -- (398.5,258) -- cycle ;
\draw  [draw opacity=0][fill={rgb, 255:red, 255; green, 0; blue, 0 }  ,fill opacity=0.1 ] (222.06,434.87) .. controls (168.07,391.42) and (134.11,328.37) .. (134.11,258.2) .. controls (134.11,257.93) and (134.11,257.65) .. (134.11,257.37) -- (398.78,258.2) -- cycle ;
\draw  [draw opacity=0][fill={rgb, 255:red, 80; green, 227; blue, 194 }  ,fill opacity=0.74 ] (223.05,435.5) .. controls (169.07,392.05) and (135.11,329) .. (135.11,258.83) .. controls (135.11,258.55) and (135.11,258.28) .. (135.11,258) -- (399.78,258.83) -- cycle ;

\draw (486,208.4) node  [scale=0.7] [align=left]  {$s_{i}$};
\draw (450,124.4) node  [scale=0.7] [align=left]  {$s_{i-1}$};
\draw (515.17,322.82) node  [scale=0.7] [align=left]  {$s_{i+1}$};
\draw (535,154.4) node  [scale=0.7] [align=left]  {$d_{1}$};
\draw (565,220.4) node  [scale=0.7] [align=left]  {$d_{2}$};
\draw (319,104.4) node  [scale=1] [align=left]  {$\mathcal{Z}_1$};
\draw (443,382.4) node  [scale=1] [align=left]  {$\mathcal{Z}_2$};
\draw (235,306.4) node  [scale=1] [align=left]  {$\mathcal{Z}_1 \cap \mathcal{Z}_2$};

\end{tikzpicture}
\caption{Representation of the union-bound}
\label{fig:union_bound}       
\end{figure}
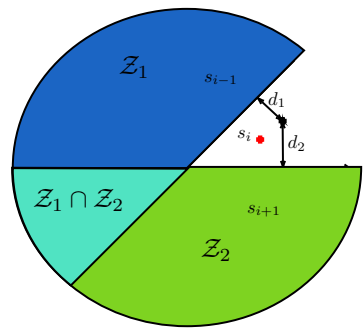

The problem of power minimization under the condition that the SEP of each user $k$ is below a given SEP requisite $p_k$ can be generally cast as 
\begin{align}
\label{opt:slp_original}
    &\min_{\boldsymbol{\theta}, P\in \mathbb{R}_+}  \ P \\
    &\hspace{0.5em}\text{s.t.}\hspace{1em} \text{P}\pc{\hat{s}_k\neq s_k|\boldsymbol{\theta},P} \leq p_k, \ \text{for} \ k\in \mathcal{K}, \notag\\
    &\hspace{1.7em}\hspace{1em} \PM{\pr{\boldsymbol{\theta}}_n}^2=1,\ \text{for} \ n \in \mathcal{N}. \notag
\end{align}
For the remainder of this study the $k$-th user correct detection probability is denoted as $\text{P}_\text{a}\pc{\hat{s}_k|\boldsymbol{\theta},P}$ and its error probability as $\text{P}_\text{e}\pc{\hat{s}_k|\boldsymbol{\theta},P}$. The detector decides for $s_k$ when the received symbol $z_k$ belongs to $\mathcal{S}_k$. With this, the probability of correct detection of the $k$-th user can be written as 
\begin{align}
    \text{P}_\text{a}\pc{\hat{s}_k|\boldsymbol{\theta},P}&=\text{P}\pc{z_k \in \mathcal{S}_k|\boldsymbol{\theta},P}=\frac{1}{\pi \sigma_w^2} \int_{\mathcal{S}_k} \hspace{-0.4em} \text{e}^{-\frac{\PM{r-\sqrt{P}\boldsymbol{h}_{k}^H\boldsymbol{\theta}}^2}{\sigma_w^2}} d{r}. \notag
\end{align}
The integral has tabled solutions for $\alpha_s\in \chav{2,4}$, yet, for $\alpha_s\notin \chav{2,4}$, solving it requires Monte Carlo methods. To achieve, for any $\alpha_s$-PSK modulation, a closed form constraint this study considers substituting $\text{P}_\text{e}\pc{\hat{s}_k|\boldsymbol{\theta},P}$ in \eqref{opt:slp_original} by the union-bound SEP. The union-bound SEP, denoted by $\text{P}_\text{ub}\pc{\hat{s}_k|\boldsymbol{\theta},P}$, is an upper-bound on the SEP \cite{lopes_tcom2023}, which implies $\text{P}_\text{e}\pc{\hat{s}_k|\boldsymbol{\theta},P}\leq \text{P}_\text{ub}\pc{\hat{s}_k|\boldsymbol{\theta},P}$. With this, meeting $\text{P}_\text{ub}\pc{\hat{s}_k|\boldsymbol{\theta},P}\leq p_k$ guarantees attainment to the original constraint. 
The union-bound states that for any finite set of events, $\text{P} \left(\bigcup _{i}A_{i}\right)\leq \sum _{i}{\text{P} }(A_{i})$, with $A_i$ representing an event.
With this, $\text{P}_\text{e}\pc{\hat{s}_k|\boldsymbol{\theta},P}$ is bounded by 
\begin{align}
\label{eq:union_bound}
        \text{P}_\text{e}\pc{\hat{s}_k|\boldsymbol{\theta},P}& =\text{P}\pc{z_k \in \mathcal{Z}_1 \cup\mathcal{Z}_2 |\boldsymbol{\theta},P} \\
        &\leq \text{P}\pc{z_k \in \mathcal{Z}_1|\boldsymbol{\theta},P} + \text{P}\pc{z_k \in \mathcal{Z}_2|\boldsymbol{\theta},P} \notag\\ 
        &= \text{P}_\text{ub}\pc{\hat{s}_k|\boldsymbol{\theta},P}, \notag
\end{align}
with $\mathcal{Z}_1$ and $\mathcal{Z}_2$ depicted in Fig.~\ref{fig:union_bound}.
The individual probabilities are computed based on the MDDTs, $d_{1,k}$ and $d_{2,k}$, as 
\begin{align}
    \text{P}\pc{z_k \in \mathcal{Z}_1|\boldsymbol{\theta},P}&=\int_{d_{1,k}}^\infty \frac{1}{\sqrt{\pi \sigma_w^2}}\  e^{-\frac{t^2}{\sigma_w^2}} dt=\frac{1}{2}\text{erfc}\pc{\frac{d_{1,k}}{\sigma_w}}\notag\\
    \text{P}\pc{z_k \in \mathcal{Z}_2|\boldsymbol{\theta},P}&=\int_{d_{2,k}}^\infty \frac{1}{\sqrt{\pi \sigma_w^2}}\  e^{-\frac{t^2}{\sigma_w^2}} dt
    =\frac{1}{2}\text{erfc}\pc{\frac{  d_{2,k}}{\sigma_w}}.\notag
\end{align}
The MDDTs are computed by applying a rotation to the coordinate system such that the symbol of interest is placed on the real axis. 
This is done by multiplying both $s_k$ and $\boldsymbol{h}_k^H \boldsymbol{\theta}$ by $s_k^*$ which results in $s_k^* s_k=1$ and $\omega_k= s_k^* \boldsymbol{h}_k^H \boldsymbol{\theta}$.
With the rotated coordinate system the MDDTs are computed as
\begin{align}
    d_{1,k}&=\sqrt{{P}}\pc{\text{Re}\{s_k^*\boldsymbol{h}_k^H \boldsymbol{\theta}\} \sin{\phi}-\text{Im}\{s_k^* \boldsymbol{h}_k^H \boldsymbol{\theta}\} \cos{\phi}}\notag\\
    d_{2,k}&=\sqrt{P}\pc{\text{Re}\{s_k^* \boldsymbol{h}_k^H \boldsymbol{\theta}\} \sin{\phi}+\text{Im}\{s_k^*\boldsymbol{h}_k^H \boldsymbol{\theta}\} \cos{\phi}}.\notag
\end{align}
With this, the bound on $\text{P}_\text{e}\pc{\hat{s}_k|\boldsymbol{\theta},P}$ is given by
\begin{align}
\label{eq:individual_error_probability}
    {\text{P}_\text{e}}\pc{\hat{s}_k|\boldsymbol{\theta},P}& \leq\text{P}_\text{ub}\pc{\hat{s}_k|\boldsymbol{\theta},P}\\
    &=\frac{1}{2}\text{erfc}\pc{\frac{d_{1,k}\pc{\boldsymbol{\theta},P}}{\sigma_w}} +\frac{1}{2}\text{erfc}\pc{\frac{d_{2,k}\pc{\boldsymbol{\theta},P}}{\sigma_w}}. \notag
\end{align}
Substituting \eqref{eq:individual_error_probability} in \eqref{opt:slp_original} yields
\begin{align}
\label{opt:complex_form}
    &\min_{\boldsymbol{\theta}, P}  \ P \\
    &\hspace{0.7em}\text{s.t.}\hspace{0.5em} \PM{\pr{\boldsymbol{\theta}}_n}^2=1,\ \text{for} \ n \in \mathcal{N}, \hspace{1em} P\geq 0, \notag \\
    &\hspace{0.7em} \sum_{\nu=1}^2 \frac{1}{2}\text{erfc}\pc{\frac{d_{\nu,k}\pc{\boldsymbol{\theta},P}}{\sigma_w}} \leq p_k, \ \text{for} \ k\in \mathcal{K}.\notag 
\end{align}
The corresponding real-valued formulation of \eqref{opt:complex_form} is cast as 
\begin{align}
\label{opt:rv_slp_unconstrained}
    &\min_{\boldsymbol{\Theta}\in \mathcal{M},P\in \mathbb{R}_+}  \ P \\
   &\hspace{0.5em} \displaystyle \sum_{\nu=1}^2 \frac{1}{2}\text{erfc}\pc{ {\sqrt{\frac{P}{\sigma_w^2}}}\ \text{tr}\pc{\boldsymbol{\Theta}  \boldsymbol{U}_{\nu,k} } } \leq p_k, \ \text{for} \ k\in \mathcal{K}. \notag
\end{align}
where,
\begin{align}
\mathcal{M}&=\{\boldsymbol{\Theta} \in \mathbb{R}^{2\times N}:  [\boldsymbol{\Theta}^T\boldsymbol{\Theta}]_{(n,n)}=1, \ \text{for} \ n\in \mathcal{N}\},\\
    \boldsymbol{U}_{1,k}&=\begin{bmatrix}
        \text{Re}\{\boldsymbol{a}^T_k\} \sin(\phi) - \text{Im}\{\boldsymbol{a}^T_k\} \cos(\phi)\\
        -\text{Re}\{\boldsymbol{a}^T_k\} \cos(\phi) - \text{Im}\{\boldsymbol{a}^T_k\} \sin(\phi)
    \end{bmatrix}^T,\\[5pt]
    \boldsymbol{U}_{2,k}&=\begin{bmatrix}
        \text{Re}\{\boldsymbol{a}_k^T\} \sin(\phi) + \text{Im}\{\boldsymbol{a}_k^T\} \cos(\phi)\\
        \text{Re}\{\boldsymbol{a}_k^T\} \cos(\phi) - \text{Im}\{\boldsymbol{a}_k^T\} \sin(\phi)
    \end{bmatrix}^T,\\[5pt]    
    \boldsymbol{\Theta}&=\begin{bmatrix}
        \pc{\text{Re}\{\boldsymbol{\theta}\}}^T \\
        \pc{\text{Im}\{\boldsymbol{\theta}\}}^T
    \end{bmatrix},
    \quad 
    \boldsymbol{a}_k=s_k^*\boldsymbol{h}_k^H.
\end{align}

\section{Solution via the Bisection Method}
\label{sec:bisection}
For solving \eqref{opt:rv_slp_unconstrained} this study considers the utilization of the bisection method (BM). The method is initialized with $P_-$ as a lower bound on $P_\text{opt}$ and $P_+$ as an upper bound on $P_\text{opt}$. The variable $P$ is fixed as $P_0=\sfrac{(P_+ + P_-)}{2}$ and the remaining problem's feasibility is evaluated. If feasible, $P_+$ is updated as $P_0$, otherwise, $P_-$ is updated as $P_0$. This is done recursively until the power difference between two consecutive iterations is below an optimality tolerance $\epsilon_\text{tol}$. For a given $P$, \eqref{opt:rv_slp_unconstrained} is rewritten as 
\begin{align}
\label{opt:sub}
    &\displaystyle \text{find}_{\boldsymbol{\Theta} \in \mathcal{M}}\ \ \boldsymbol{\Theta} \\
    &\hspace{0.5em}\text{s.t.}\displaystyle \sum_{\nu=1}^2 \frac{1}{2}\text{erfc}\pc{ {\sqrt{\frac{P}{\sigma_w^2}}}\ \text{tr}\pc{\boldsymbol{\Theta}  \boldsymbol{U}_{\nu,k} } } \leq p_k, \ \text{for} \ k\in \mathcal{K}. \notag
\end{align}
The strategy for solving \eqref{opt:sub} consists of minimizing the maximum constraint and evaluating if the optimal solution attains it. This yields the following optimization problem
\begin{align}
\label{opt:subproblem2}
    \boldsymbol{\Theta}_\text{opt}&=\displaystyle \min_{\boldsymbol{\Theta \in \mathcal{M}}}\ \max_{k\in \mathcal{K}}\pc{\displaystyle \sum_{\nu=1}^2 \frac{1}{2}\text{erfc}\pc{ {\sqrt{\frac{P}{\sigma_w^2}}}\ \text{tr}\pc{\boldsymbol{\Theta}  \boldsymbol{U}_{\nu,k} } } - p_k}. 
\end{align}
Based on $\boldsymbol{\Theta}_\text{opt}$ the feasibility of \eqref{opt:subproblem2} is evaluated by checking $f(\boldsymbol{\Theta}_\text{opt})\leq 0$, where
\begin{align}
\label{eq:f}
    f(\boldsymbol{\Theta})=\max_{k\in \mathcal{K}}\pc{\sum_{\nu=1}^2 \frac{1}{2}\text{erfc}\pc{ {\sqrt{\frac{P}{\sigma_w^2}}}\ \text{tr}\pc{\boldsymbol{\Theta} \boldsymbol{U}_{\nu,k} } } - p_k}.
\end{align}
If the condition holds the problem is feasible and $\boldsymbol{\Theta}_\text{opt}$ is a solution of \eqref{opt:sub}. Otherwise, there is at least one constraint that cannot be fulfilled with the given transmit power $P$, implying that \eqref{opt:sub} is infeasible. The steps of the BM are further detailed in algorithm \ref{alg:bisection}.

\begin{algorithm}
\small
  \caption{Proposed Bisection Method for solving \eqref{opt:rv_slp_unconstrained}}
	\label{alg:bisection}
  \begin{algorithmic}    
  \State{\textbf{Inputs}: ${P}_+ \geq P_\text{opt}$, ${P}_-\leq P_\text{opt}$, $f_a<0$, $i_\text{max}>0$, $\epsilon_\text{tol}>0$ \hspace{1em} \textbf{Output}: $P_\text{opt}$, $\boldsymbol{\theta}_\text{opt}$ }
  \State{Define $i=0$, $P_a=P_0$}
  \State{\textbf{While} $\pc{ {P_{a}-P_{0}} \leq \epsilon_\text{tol}\ \cup \ i\leq i_\text{max}} \ \cap \ f_a< 0$}
  \State{\hspace{0.5em} Solve \eqref{opt:subproblem3} with RCG \cite{boumal_manopt} considering $P=P_0$ and get $\boldsymbol{\Theta}_\text{opt}$}
  \State{\hspace{0.5em} Compute $f_a=f(\boldsymbol{\Theta}_\text{opt})$ with \eqref{eq:f}}
  \State{\hspace{0.5em} \textbf{If} $f_a\leq0$ $\rightarrow$ Update $P_{+}={P_0}$}
  \State{\hspace{0.5em} \textbf{Else} Update $P_{-}={P_0}$}
 \State{\hspace{0.5em} Update $P_a=P_0$, $P_0=\sfrac{\pc{P_+ + P_-}}{2}$ and $i={i+1}$}
\State{ Update $P_\text{opt}=P_0$ and $\boldsymbol{\theta}_\text{opt}=\pr{\boldsymbol{\Theta}_\text{opt}}_{(1,:)}+j\pr{\boldsymbol{\Theta}_\text{opt}}_{(2,:)}$}
    \end{algorithmic}
\end{algorithm}

As mentioned in \cite{lopes_tcom2023}, the function $\sum_i\text{erfc}\pc{\cdot}$ is convex if the argument is nonnegative. This implies that $f(\boldsymbol{\Theta})$ is convex for $\text{\text{tr}\pc{\boldsymbol{\Theta} \boldsymbol{U}_{\nu,k} }}\geq0$, for $k\in \mathcal{K}$ and $\nu\in\{1,2\}$, or, equivalently, for $d_{\nu,k}(\boldsymbol{\theta},P) \geq0$, for $k\in \mathcal{K}$ and $\nu\in\{1,2\}$. This in turn makes problem \eqref{opt:subproblem2} non-convex. Yet, as presented in what follows, due to the characteristics of the problem it can be solved with descent algorithms if a suitable initialization point is provided.

\subsection{Evaluating Feasibility via Riemannian Conjugate Gradient}

The utilization of algorithm \ref{alg:bisection} implies a method for solving the unconstrained problem in \eqref{opt:subproblem2}. In this study this is done with the RCG algorithm \cite[Section 3.1]{boumal2014optimization}, which is designed for solving unconstrained minimization problems in Riemannian manifolds.
Since the RCG approach requires a twice continuously differentiable objective, $f(\boldsymbol{\Theta})$ is substituted by its softmax approximation computed with the log-sum-exp function $\text{LSE}(\boldsymbol{x})=\text{ln}\pc{\sum_{i} e^{x_i}}$, which yields
\begin{align}
\label{eq:f0}
    f_0 (\boldsymbol{\Theta})=\text{ln}\pc{ \sum_{k=1}^K e^{\sum_{\nu=1}^2 \frac{1}{2}\text{erfc}\pc{ {\sqrt{\frac{P}{\sigma_w^2}}}\ \text{tr}\pc{\boldsymbol{\Theta}  \boldsymbol{U}_{\nu,k} } } - p_k}}.
\end{align}
The precision of the approximation obeys the following bound
\begin{align}
    \displaystyle \max_{\{i=1,\hdots,n\}} {x_i}\leq \mathrm {LSE} (x_{1},\dots ,x_{n})\leq \max_{\{i=1,\hdots,n\}} {x_i}+\log(n). \notag
\end{align}
Note that, $\text{LSE}(\boldsymbol{x})$ is a nondecreasing function, which implies that for the regions where $f(\boldsymbol{\Theta})$ is convex the convexity is preserved. With this, \eqref{opt:subproblem2} is rewritten as
\begin{align}
\label{opt:subproblem3}
    &\displaystyle \min_{\boldsymbol{\Theta \in \mathcal{M}}}\ \text{ln}\pc{ \sum_{k=1}^K e^{\sum_{\nu=1}^2 \frac{1}{2}\text{erfc}\pc{ {\sqrt{\frac{P}{\sigma_w^2}}}\ \text{tr}\pc{\boldsymbol{\Theta}  \boldsymbol{U}_{\nu,k} } } - p_k}} 
\end{align}
For computing the solution, the RCG algorithm utilizes the Euclidean gradient of the objective $f_0(\boldsymbol{\Theta})$ which is given by
\begin{align}
    &\nabla f_0(\boldsymbol{\Theta})=\pc{\sum_{k=1}^K{e^{h_k(\boldsymbol{\Theta})}\nabla h_k}}\pc{\sum_{k=1}^K e^{h_k(\boldsymbol{\Theta})}}^{-1} \notag\\
    &h_k(\boldsymbol{\Theta})=\sum_{\nu=1}^2 \frac{1}{2}\text{erfc}\pc{ {\sqrt{\frac{P}{\sigma_w^2}}}\ \text{tr}\pc{\boldsymbol{\Theta}  \boldsymbol{U}_{\nu,k} } } - p_k, \ \text{for}\ k\in\mathcal{K}, \notag\\
    &\nabla h_k(\boldsymbol{\Theta})=- \sqrt{\frac{P}{\pi\sigma_w^2}}\sum_{\nu=1}^2 e^{- {\frac{P\cdot \text{tr}\pc{\boldsymbol{\Theta}  \boldsymbol{U}_{\nu,k} }^2}{\sigma_w^2}} } \boldsymbol{U}_{\nu,k}^T, \ \text{for}\ k\in\mathcal{K}.\notag
\end{align}

Similar to other descent algorithms the RCG requires initialization with a strictly feasible point. If the objective function $f(\boldsymbol{\Theta})$, and consequently $f_0(\boldsymbol{\Theta})$,
would be convex for $\boldsymbol{\Theta}\in \mathbb{R}^{2\times N}$ any starting point $\boldsymbol{\Theta}\in \mathcal{M}$ would lead to a local optimal solution. Yet, the function $f(\boldsymbol{\Theta})$ is convex for $\boldsymbol{\Theta}\in \mathcal{Y}$ with $\mathcal{Y}=\{\boldsymbol{\Theta}: \text{tr}\pc{\boldsymbol{\Theta} \boldsymbol{U}_{\nu,k}}\geq0,\  \text{for}\ k \in \mathcal{K}, \ \nu \in\{1,2\}\}$, 
which implies convexity of $f_0(\boldsymbol{\Theta})$ for $\boldsymbol{\Theta}\in \mathcal{Y}$.

Due to $f_0(\boldsymbol{\Theta})$ being convex for $\boldsymbol{\Theta}\in \mathcal{Y}$, if the optimal solution of \eqref{opt:subproblem3}, termed $\boldsymbol{\Theta}_\text{opt}$, belongs to $\mathcal{Y}$, initializing the RCG method with any value of $\boldsymbol{\Theta}_0 \in\mathcal{Y}$ supports finding a local optimal solution. This is the case since $f_0(\boldsymbol{\Theta})$ grows for a decrease in $\text{tr}\pc{\boldsymbol{\Theta} \boldsymbol{U}_{\nu,k}}$ and thus by initializing the RCG algorithm with $\boldsymbol{\Theta}_0 \in\mathcal{Y}$ it will takes steps to stay on $\mathcal{Y}$.

On the other hand, if $\boldsymbol{\Theta}_\text{opt}\notin \mathcal{Y}$, which corresponds to a noiseless received signal outside the correct decision region, the value of at least one constraint function $h_k(\boldsymbol{\Theta})$ is given by $h_k(\boldsymbol{\Theta})\geq 0.5-p_k$. Since $f(\boldsymbol{\Theta})=\max_{k} h_k(\boldsymbol{\Theta})$, this implies that in this case \eqref{opt:rv_slp_unconstrained} is infeasible for $p_k<0.5 \ \forall k\in\mathcal{K}$. For this case, initializing the RCG algorithm for solving \eqref{opt:subproblem3} with any starting point, including $\boldsymbol{\Theta}_0 \in\mathcal{Y}$, will yield an output $\boldsymbol{\Theta}_\text{out}$ such that $f(\boldsymbol{\Theta}_\text{out})>0$.
With this, the optimization problem for computing the initial point $\boldsymbol{\Theta}_0\in \mathcal{Y}$ can be cast as 
\begin{align}
\label{opt:trace}
    &\displaystyle \max_{\boldsymbol{\Theta \in \mathcal{M}}}\ \min_{k \in \mathcal{K}, \nu \in \{1,2\}} \text{tr} \pc{\boldsymbol{\Theta}  \boldsymbol{U}_{\nu,k} }.
\end{align}
To solve \eqref{opt:trace} via the RCG algorithm the log-sum-exp function is applied which yields the following problem
\begin{align}
\label{opt:initial_problem}
    &\displaystyle \boldsymbol{\Theta}_0=\min_{\boldsymbol{\Theta \in \mathcal{M}}}\ v_0(\boldsymbol{\Theta}),
\end{align}
with $v_0(\boldsymbol{\Theta})=\text{ln}\pc{ \sum_{k=1}^K e^{-\text{tr}\pc{\boldsymbol{\Theta}  \boldsymbol{U}_{1,k} }} + e^{-\text{tr}\pc{\boldsymbol{\Theta}  \boldsymbol{U}_{2,k} }}}$, and,
\begin{align}
\label{eq:grad_initial}
    \nabla v_0(\boldsymbol{\Theta})=- \frac{\sum_{k=1}^K e^{-\text{tr}\pc{\boldsymbol{\Theta}  \boldsymbol{U}_{1,k} }}\boldsymbol{U}_{1,k}^T+e^{-\text{tr}\pc{\boldsymbol{\Theta}  \boldsymbol{U}_{2,k} }}\boldsymbol{U}_{2,k}^T}
    {\sum_{k=1}^K e^{-\text{tr}\pc{\boldsymbol{\Theta}  \boldsymbol{U}_{1,k} }} + e^{-\text{tr}\pc{\boldsymbol{\Theta}  \boldsymbol{U}_{2,k} }}}.
\end{align}
Problem \eqref{opt:initial_problem} is solved via the utilization of the RCG algorithm with any starting point $\boldsymbol{\Theta}\in \mathcal{M}$. The details of the RCG implementation are given in \cite{boumal_manopt}.

\subsection{Computational Complexity}

As mentioned in \cite{Joint_SLP_IRS} the upper bound complexity for executing the RCG algorithm is in the order of $\mathcal{O}\pc{N^{1.5}}$. Since the number of times that RCG is required to run mainly depends on initialization of $P_+$ and $P_-$ and does not grow with the size of the system the overall upper bound complexity of the proposed algorithm is in the order of $\mathcal{O}\pc{N^{1.5}}$.

\section{Numerical Results}
\label{sec:numerical_results}

\begin{figure}[t]
\begin{center}
%
%
%
\usetikzlibrary{positioning,calc}

\definecolor{mycolor1}{rgb}{0.00000,1.00000,1.00000}%
\definecolor{mycolor2}{rgb}{1.00000,0.00000,1.00000}%

\pgfplotsset{every axis label/.append style={font=\footnotesize},
every tick label/.append style={font=\footnotesize}
}

\begin{tikzpicture}[spy using outlines={rectangle,magnification=3,connect spies}] 
\begin{axis}[%
name=A,
ymode=linear,
width  = 0.80\columnwidth,
height = 0.45\columnwidth,
scale only axis,
xmin  = 1,
xmax  = 10,
xlabel= {$\tau$},
xmajorgrids,
ymin=-23,
ymax=-3,
ylabel={$P_n$ [dB]},
ymajorgrids,
legend entries={
},
                legend columns=1,
legend style={at={(0,1)},anchor=north west ,draw=black,fill=white,legend cell align=left,font=\tiny}
]




\addplot+[smooth,color=dark_green,solid, every mark/.append style={solid, fill=cyan!50},mark=pentagon,
y filter/.code={\pgfmathparse{\pgfmathresult-0}\pgfmathresult}]
  table[row sep=crcr]{%
1        -15.7358    \\
2        -11.7870    \\
3         -9.5732    \\
4         -8.0559    \\
5         -6.9049    \\
6         -6.0059    \\
7         -5.2251    \\
8         -4.5622    \\
9         -3.9865    \\
10        -3.4789    \\
};

\addplot+[smooth,color=purple,solid, every mark/.append style={solid, fill=cyan!50},mark=o,
y filter/.code={\pgfmathparse{\pgfmathresult-0}\pgfmathresult}]
  table[row sep=crcr]{%
1       -18.5818    \\
2        -14.6509    \\
3        -12.4815    \\
4        -10.9848    \\
5         -9.8479    \\
6         -8.9350    \\
7         -8.1691    \\
8         -7.5148    \\
9         -6.9402    \\
10        -6.4291    \\
};

\addplot+[smooth,color=red,solid, every mark/.append style={solid, fill=cyan!50},mark=square,
y filter/.code={\pgfmathparse{\pgfmathresult-0}\pgfmathresult}]
  table[row sep=crcr]{%
1          -20.5909   \\
2          -16.6781   \\
3          -14.5213   \\
4          -13.0378   \\
5          -11.9107    \\
6          -10.9987   \\
7          -10.2409  \\
8           -9.5296  \\
9           -9.0234  \\
10          -8.5138  \\
};

\addplot+[smooth,color=blue,solid, every mark/.append style={solid, fill=cyan!50},mark=x,
y filter/.code={\pgfmathparse{\pgfmathresult-0}\pgfmathresult}]
  table[row sep=crcr]{%
1           -22.5871  \\
2           -18.6744  \\
3           -16.5319  \\
4           -15.0603  \\
5           -13.9393   \\
6           -13.0386  \\
7           -12.2855 \\
8           -11.6356 \\
9           -11.0695 \\
10          -10.5668 \\
};

\addplot[smooth,color=red,mark=square,solid,
y filter/.code={\pgfmathparse{\pgfmathresult-100}\pgfmathresult}]
  table[row sep=crcr]{%
	1 2\\
};\label{plot:sim_sep}

\addplot[smooth,color=dark_green,mark=pentagon,solid,
y filter/.code={\pgfmathparse{\pgfmathresult-100}\pgfmathresult}]
  table[row sep=crcr]{%
	1 2\\
};\label{plot:exact_sep}

\addplot[smooth,color=purple,mark=o,solid,
y filter/.code={\pgfmathparse{\pgfmathresult-100}\pgfmathresult}]
  table[row sep=crcr]{%
	1 2\\
};\label{plot:ub_sep}

\addplot[smooth,color=blue,mark=x,solid,
y filter/.code={\pgfmathparse{\pgfmathresult-100}\pgfmathresult}]
  table[row sep=crcr]{%
	1 2\\
};\label{plot:sixty}

\node [draw,fill=white,font=\tiny,anchor= south  east] at (axis cs: 10,-23) {
\setlength{\tabcolsep}{0.5mm}
\renewcommand{\arraystretch}{.8}
\begin{tabular}{l}
\ref{plot:exact_sep}{\hspace{0.5em} $N=30$}\\
\ref{plot:ub_sep}{\hspace{0.5em} $N=40$}\\
\ref{plot:sim_sep}{\hspace{0.5em} $N=50$} \\
\ref{plot:sixty}{\hspace{0.5em} $N=60$} \\
\end{tabular}
};

\end{axis}

\end{tikzpicture}%
\caption{Normalized Transmit Power [dB] $\times$ $\tau$, for $K=5$, $\alpha_s=4$ and $p_k=10^{-\tau}$ for $k \in \mathcal{K}$.} 
\label{fig:power}       
\end{center}
\end{figure}
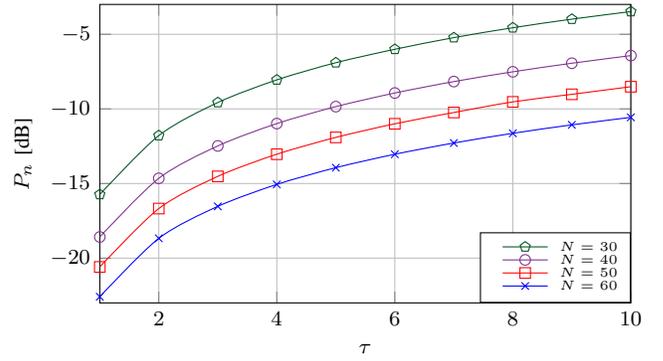

In this section, the proposed method is numerically evaluated considering a system with $K=5$ users and QPSK data, i.e., $\alpha_s=4$, for a RIS with $N=30$, $N=40$, $N=50$ and $N=60$ reflecting elements. For the simulation it was considered $\sigma_w^2=1$, $P_-=0$, $P_+=100$, $\epsilon_\text{tol}=10^{-7}$, and $2\cdot10^3$ symbols were transmitted. To facilitate the analysis all users are considered to have the same SEP requirement such that ${p}_k=10^{-\tau}$, for $k\in\mathcal{K}$. The method is evaluated in terms of average normalized transmit power $P_n=10 \log_{10}(\sfrac{P}{\sigma_w^2})$ [dB] required for attaining the SEP constraints versus $\tau$. Fig.~\ref{fig:power} shows that the proposed power minimization SLP achieves the SEP requisite with reduced average transmit power. Fig.~\ref{fig:power} also shows that there is a significant decrease in average transmit power for every increase in the number of reflecting elements. 



\section{Conclusion}
\label{sec:conclusion}

This study considers a virtual MU-MIMO system realized with the RIS-based passive transmitter setup. For the considered framework the study derives the formulation for the union-bound SEP and proposes a SLP power minimization problem under the condition that the union-bound SEP is below a given requirement.
The problem is formulated as a constrained optimization on an oblique manifold, and solved via the application of a BM to successively optimize transmit power while evaluating the feasibility of the union-bound SEP requisite by solving an auxiliary problem dependent only on the RIS reflection coefficients with RCG. Numerical results show that the proposed approach yields reduced normalized transmit power for different SEP requirements.

%
\newpage
\bibliographystyle{IEEEtran}
\bibliography{bib-refs}

\end{document}